\documentclass[
twocolumn,
showpacs, amssymb, aps,prl]{revtex4}

\usepackage{graphicx}
\usepackage{bm}

\begin{document}

\title{Zener Tunneling Between Landau Orbits\\ in a High-Mobility Two-Dimensional
         Electron Gas}
\author{C. L. Yang}
\author{J. Zhang}
\author{R. R. Du}
\affiliation{Department of Physics, University of Utah, Salt Lake City, Utah 84112}
\author{J. A. Simmons}
\author{J. L. Reno}
\affiliation{Sandia National Laboratories, Albuquerque, New Mexico 87185}
\begin{abstract}
Magnetotransport in a laterally confined two-dimensional electron gas (2DEG) can 
exhibit modified scattering channels owing to a tilted Hall potential. 
Transitions of electrons between Landau levels with shifted guiding centers
can be accomplished through a Zener tunneling mechanism, and make a 
significant contribution to the magnetoresistance. A remarkable oscillation 
effect in weak field magnetoresistance has been observed in high-mobility 
2DEGs  in GaAs-AlGa$_{0.3}$As$_{0.7}$ heterostructures, and can be well
explained by the Zener mechanism.
\end{abstract}


\maketitle

Scattering and dissipation are central issues in quantum transport in
electronic systems. Of particular interest are the peculiar phenomena in a
quantum Hall (QH) system, realized when a two-dimensional electron gas (2DEG)
is subject to a strong magnetic field, $B$. In the integer quantum Hall effect
(IQHE) regime the Hall plateau is formed and the longitudinal transport is
dissipationless~\cite{Prange1990}. It has been known for quite sometime that, 
when a substantial current is passed through the 2DEG, the IQHE tends to break
down. Above a certain critical current density the quantized Hall plateau disappears 
and the transport in this regime becomes dissipative. Such phenomena have been 
observed in confined QH systems, such as Hall bar~\cite{Ebert_Cage1983} or Corbino 
geometries~\cite{Simon1986}. 

The mechanism leading to the breakdown effect is still under debate. Among
the possible explanations, the Zener tunneling mechanism was originally proposed 
by Tsui \textit{et al}~\cite{Tsui1983}. In essence, if a sufficient Hall field, $E_{y}$, 
is established in a laterally confined QH system, the degeneracy of the Landau orbits
is lifted. Zener tunneling can occur between the occupied Landau orbits below the 
Fermi level, $E_{F}$, and the empty orbits above the $E_{F}$, separated at a 
distance equivalent to the cyclotron diameter, $2R_{c}$. The condition for energy 
conservation is satisfied whenever $2R_{c}eE_{y}=l\hbar\omega_{c}$, where 
$l=1, 2, 3 ...$ integers, $\omega_{c}=eB/m^{\ast}$ the cyclotron frequency, 
and $m^{\ast}$ the electron effective mass. By this mechanism, a critical current density 
of $j\sim 60\ A/m$ is needed to initiate the IQHE breakdown in GaAs, at a typical magnetic field 
of $B\sim10\ T$~\cite{Tsui1983}. In the breakdown effect of IQHE, such 
tunneling events are likely to take place near the edges of the QH system where a high 
field can build up.

Quite surprisingly, we have observed the Zener tunneling effect in a
different regime. Specifically, our effect takes place in the bulk of a 2DEG
subjected to a weak magnetic field. Remarkable magnetoresistance oscillations 
are observed in high-mobility Hall bar specimens. The period of oscillation, $\Delta(1/B)$, 
is tunable by a dc bias current, $J_{dc}$: $\Delta(1/B)\propto n_{e}^{1/2}/J_{dc}$, 
where $n_{e}$ is the electron sheet density. This observation confirms a general selection 
rule for electron transport in a weak magnetic field, namely, $\Delta k_{x} \approx 2k_{F}$ for 
momentum transfer between the initial and final Landau levels, where 
$x$ labels the direction of the current  and $2k_{F}$ is the Fermi wave vector of the 2DEG
at zero magnetic field. This selection rule in momentum space translates into 
$\Delta Y \approx 2R_{c}$ in real space, where $\Delta Y$ is the guiding center shift. Such a 
modulation of the scattering phase space is responsible for an oscillatory magnetoresistance, 
equivalent to a geometrical resonance~\cite{Gerhardts_Winkler1989} without a potential  
modulation. 

We report in this letter the observations for the new oscillations, and propose 
a simple  model which can account for all the major features. 
The roles played by the short-range scatterers in this effect, and possible 
experiments to explore Zener tunneling between composite fermion (CF) orbits 
in a half-filled Landau level~\cite{Sarma1997} will be briefly discussed.   

Our samples were cleaved from a wafer of a high-mobility 
GaAs-Al$_{0.3}$Ga$_{0.7}$As heterostructure grown by molecular-beam 
epitaxy, having an electron density $n_{e}\approx2$ $\times10^{11}\ cm^{-2}$ 
and a mobility $\mu\approx3\times10^{6}\ cm^{2}/{Vs}$ at a temperature 
$T=4.2\ K$. Such parameters were obtained after a brief illumination from 
a light-emitting diode. The distance between the electrons and the 
Si $\delta$-doping layer is $d_{s}\approx70\ nm$. Four Hall bar specimens of 
width $w=200,\ 100,\ 50$, and $20\ \mu m$ were processed by photolithography 
and wet etching. The 50 $\mu m$ specimen has a NiCr front gate so that its 
electron density can be tuned between 1.9 and $4.0\times10^{11}\ cm^{-2}$. 
The experiments were performed in a sorption-pumped $^{3}He$ cryostat 
equipped with a superconducting magnet. 

In principle, the Zener tunneling effect can be detected in standard
magnetoresistance, $R_{xx}(B)$. However, in order to increase the sensitivity,
a differential resistance $r_{xx}$ was measured in the following fashion. A
constant dc current $I_{dc}$ was passed through the Hall bar, along with a
small ($100\ nA$) low frequency ($f=23\ Hz$) modulation current, 
$i_{ac}$. The differential magnetoresistance at the given dc bias,  
$r_{xx}=(\partial V/\partial I)_{I_{dc}}=v_{ac}/i_{ac}$, was then 
recorded by a lock-in amplifier at the modulation frequency. A schematic circuit 
for the electrical measurement is shown in the inset of Fig. 1.

\begin{figure}
\includegraphics{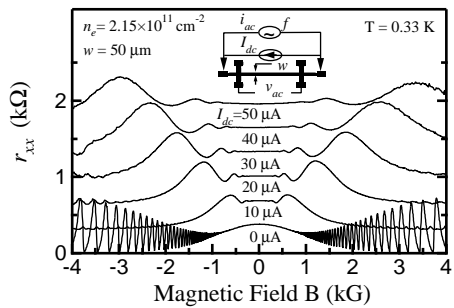}
\caption{\label{fig1} The measured differential magnetoresistance traces at 
various dc current $I_{dc}$ are shown for a 50 $\mu m$ Hall bar (for clarity, 
the traces are shifted vertically in steps of $0.3\ k\Omega$). Up to three orders 
of oscillations are clearly seen from the traces, and the oscillations are roughly 
periodic in $1/B$. The inset is a diagram for the electrical measurement.}
\end{figure}

Our central finding concerns the strong oscillation in $r_{xx}$ in the weak magnetic 
field region, $B<4\ kG$. In Fig.~1 we show such features from a 50 $\mu m$ Hall 
bar measured at $T=0.33\ K$, for $I_{dc}=0,\ 10,\ 20,\ 30,\ 40,\ 50\ \mu A$, 
respectively. For a zero dc bias current, the trace shows well-resolved Shubnikov 
de-Haas (SdH) oscillations for $B>0.5\ kG$. New oscillations emerge when a 
finite $I_{dc}$\ is applied to the specimen. Up to three orders of peaks can be 
clearly seen from the traces. Furthermore, the peaks shift towards higher $B$ with 
increasing $I_{dc}$. The weakening of the SdH effect can be attributed partially to 
electronic heating by the dc bias. In contrast to the SdH oscillation whose amplitude 
diminishes quickly at increasing temperatures, the new oscillations persist to 
a temperature up to $T=4K$.

It is apparent that the new oscillations, observed here in $r_{xx}$, are roughly 
periodic in $1/B$. However, as can be shown,  the exact resonance condition 
for Zener tunneling should correspond to peaks in its derivative, \textit{viz.}, 
in $\partial r_{xx}/\partial|B|$ traces~\cite{data_analysis}. To illustrate this point, 
we show in Fig.~2(a) the $\partial r_{xx}/\partial|B|$\ trace which is obtained by 
numerical differentiation performed on the $r_{xx}(B)$ trace from a 50 $\mu m$ 
sample with a bias $I_{dc}=30\ \mu A$. The inset, $1/B_{l}$ vs. $l$,
confirms that the oscillation in $\partial r_{xx}/\partial|B|$ is strictly periodic in $1/B$. 

\begin{figure}
\includegraphics{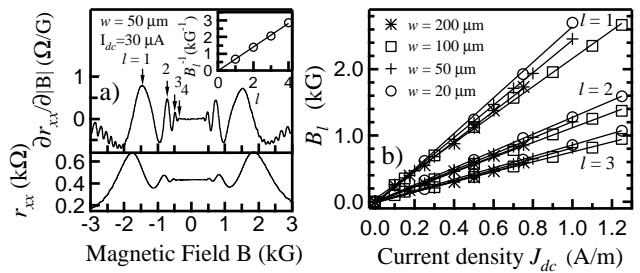}
\caption{\label{fig2} 
(a) The differential magnetoresistance $r_{xx}$\ at $I_{dc}=30\ \mu
A$\ for the 50 $\mu m$ Hall bar is illustrated together with its derivative
$\partial r_{xx}/\partial|B|$. The insert shows the precise $1/B$ periodicity
extracted from the $\partial r_{xx}/\partial|B|$  trace. (b) The oscillation
maximum $B_{l}$ ($l=1,2,3$)  versus current density $J_{dc}$ are 
shown as a fan diagram for the Hall bars with different width $w=200, 100, 50,\ and\ 20\ \mu m$
(the electron densities, in units of $10^{11}\ cm^{-2}$,  are 2.08, 1.94, 2.15\ and\ 2.14), 
respectively. A linear fit precisely reveals the relation $B_{l}\propto J_{dc}/l$, 
and the slope is nearly the same for all the samples.  }
\end{figure}

Remarkably, the oscillation period is tunable by the bias current $I_{dc}$.
In Fig.~2 (b) we plot the maximum positions $B_{l}$ (1, 2, 3), obtained from
$\partial r_{xx}/\partial|B|$, against the current density $J_{dc}=I_{dc}/w$,
for the four samples. Roughly, all data collapse according to 
\begin{equation}
  B_{l}\propto J_{dc}/l.
\label{eq:observation}
\end{equation}

Such oscillations arise owing to a new scattering channel opened up by a tilted 
Hall potential. To begin with we consider a 2DEG system under crossed electric 
and magnetic fields, depicted in Fig.~3. The electric field is a Hall field along 
the $y$ direction, $E_{y}=v_{d}B$, induced by a dc current density 
$J_{dc}=n_{e}ev_{d}$, where $v_{d}$ is the drift velocity of the electrons.

The 2DEG is quantized into a series of Landau levels and has a wave
function~\cite{Kahn1959, Calecki1977}
\begin{equation}
  |NY\rangle=L_{x}^{-1/2}e^{ik_{x}x}\phi_{N}(y-Y),
\end{equation}
where $N$ is the index of Landau levels, and  $\phi_{N}(y-Y)$ is an oscillatory
wave function centered at $Y=-l_{B}^{2}(k_{x}-m^{\ast}v_{d}/\hbar)$,
with $l_{B}=\sqrt{\hbar/eB}$ the magnetic length.
 
The energy levels are given by \begin{equation}
E_{NY}=(N+\frac{1}{2})\hbar\omega_{c}-eE_{y}Y+\frac{1}{2}m^{\ast}v_{d}^{2}.
\end{equation}
Because of  the potential of the Hall field, the degeneracy of Landau
levels with respect to the guiding center $Y$ is lifted and the Landau levels are 
tilted spatially along the $y$ direction with a slope given by $eE_{y}$. 
Since the density of electrons is largely homogenous within the sample, 
the distribution function of electrons should not depend on
$Y$~\cite{Calecki1977}, which means the 
Fermi level is also tilted along the $y$ direction with the same slope as that of 
the Landau levels~\cite{Chaubet1995}. As a result, all the states within 
one Landau level with different guiding center $Y$  are equally occupied. 

\begin{figure}
\includegraphics{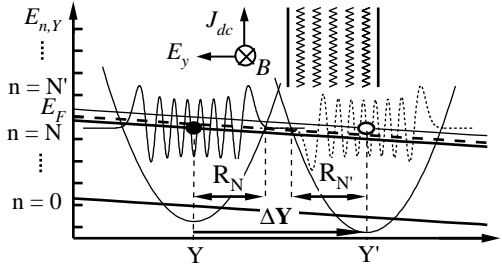}
\caption{\label{fig3} 
The Landau levels are spatially tilted along the direction of electric field 
($y$ direction), and the Fermi level has the same slope as the Landau
levels. Elastic scattering cause the electron hopping between different 
Landau levels with a distance given by $\Delta Y=Y^{\prime}-Y
=l\hbar\omega_{c}/eE_{y}$.  The vanishing of overlap between the
oscillatory wave functions when $\Delta Y>R_{N}+R_{N^{\prime}}
\approx2R_{c}$  means the maxmium hopping distance allowed 
is about $2R_{c}$. The configuration of the crossed electric and magnetic
 fields is also shown.  }
\end{figure}

In the presence of elastic scattering, an electron may transfer a momentum
$q_{x}=k_{x}-k_{x}^{\prime}$ to a scatterer, which is equivalent to a hopping 
(shifting of the guiding center) in the $y$ direction at a distance $\Delta Y=Y^{\prime
}-Y=l_{B}^{2}q_{x}$. This hopping gives a current density 
$j_{y}$~\cite{Calecki1977, Titeica1935}, hence  a conductivity
\begin{equation}
    \sigma_{yy}=\frac{j_{y}}{E_{y}}=\frac{1}{E_{y}}\frac{e}{2L_{x}L_{y}}
    \sum_{\mu\mu^{\prime}}W_{\mu\mu^{\prime}}(Y^{\prime}-Y)
    (f_{\mu}-f_{\mu^{\prime}}),
\label{sigma_yy}
\end{equation} 
where $\mu=(NY),\ \mu^{\prime}=(N^{\prime}Y^{\prime})$, $L_{x}$ and $L_{y}$
the dimensions of the 2DEG, $f_{\mu}=1/(e^{(E_{\mu}-E_{F})/k_{B}T}-1)$ 
the Fermi distribution of the electrons, and $W_{\mu\mu^{\prime}}$  the transition
rate from the initial state $|\mu\rangle$ to final state $|\mu^{\prime}\rangle$ in the 
Born approximation~\cite{Roth1966}:
\begin{eqnarray}
W_{\mu\mu^{\prime}}&=&\frac{2\pi}{\hbar} \frac{n_{i}}{L_{x}L_{y}}
    \sum_{q_{x},q_{y}}|V(q)|^{2}|\langle\mu^{\prime}|e^{i\overrightarrow{q}.
    \overrightarrow{r}}|\mu\rangle|^{2}\delta(E_{\mu}-E_{\mu^{\prime}})\nonumber\\
   &=&\frac{n_{i}}{\hbar L_{x}l_{B}} \delta(E_{\mu}-E_{\mu^{\prime}})\int
        dQ_{y}\left|V\left(\frac{Q}{l_{B}}\right)\right|^{2}J_{N}^{l}(Q),
\label{trans_rate}
\end{eqnarray}
where $Q\equiv ql_{B}=\sqrt{Q_{x}^{2}+Q_{y}^{2}},\ Q_{x}\equiv\Delta Y/l_{B},
\ n_{i}$  is the density of the random scatterers, $V(q)$ is the effective 
Fourier component of the scatterring potential seen by the 2DEG, 
and~\cite{Zudov2001}
\begin{eqnarray}
  J_{N}^{l}(Q)&=&\left|\int e^{iq_{y}y}\phi_{N}(y-Y)\phi_{N+l}(y-Y^{\prime})\right|^{2}
                         \nonumber\\
  &=&\frac{N!}{(N+l)!}\left(\frac{Q^{2}}{2}\right)^{l}
     e^{-\frac{Q^{2}}{2}}\left[L_{N}^{l}\left(\frac{Q^{2}}{2}\right)\right]^{2},
\end{eqnarray}
with $l=N^{\prime}-N$ the index difference between involved Landau levels and
$L_{N}^{l}(x)$ the generalized Laguerre polynomial.

In Eq.~(\ref{trans_rate}), the  $\delta(E_{\mu}-E_{\mu^{\prime}})$ accounts 
for the conservation of energy, which gives  $eE_{y}\Delta Y=l\hbar\omega_{c}$.
This means an electron hopping along $y$ direction should cause a transition 
between Landau levels, and the hopping distance is determined by
\begin{equation}
  \Delta Y_{l}=\frac{l\hbar\omega_{c}}{eE_{y}}=l\frac{\hbar}{m^{\ast}v_{d}}
          =l\frac{e\hbar n_{e}}{m^{\ast}J_{dc}}
\label{deltaY_fix},
\end{equation}
which does not depend on the magnetic field, and is fixed for a given current
density $J_{dc}$. The selection rule of Eq.~(\ref{deltaY_fix}) is unusual, in that 
it only stems from the two dimensional nature of the electrons. 

The resistivity along the $x$ direction (assume $\mu B\gg1$) is 
$\rho_{xx}=\sigma_{yy}/(\sigma_{xx}\sigma_{yy}+\sigma_{xy}^{2})
     \approx\rho_{xy}^{2}\sigma_{yy}$.
By working out $\sigma_{yy}$, finally we get 
\begin{equation}
\rho_{xx}=\frac{h}{e^{2}}\frac{ (2\pi)^{3}n_{i}m^{\ast 4}v_{d}^{2}}{h^{6}n_{e}^{2}}
  \sum_{Nl}(f_{N}-f_{N+l})P_{Nl}(Q_{xl}),
\label{rho_xx}
\end{equation}
with $Q_{l}\equiv\sqrt{Q_{xl}^{2}+Q_{y}^{2}},\ Q_{xl}\equiv\Delta Y_{l}/l_{B}$, and
\begin{equation}
  P_{Nl}(Q_{xl})=\frac{Q_{xl}^{5}}{l^{4}}\int dQ_{y}\left|V\left(\frac{Q_{l}Q_{xl}
    }{\Delta Y_{l}}\right)\right|^{2}J_{N}^{l}(Q_{l}).
\end{equation}

The function $P_{Nl}(Q_{xl})$ can be numerically evaluated for a given $V(q)$. 
We simply assume a constant $V(q)$, which is good for short-range scattering, 
to calculate this function. The results show that $P_{Nl}$ has a dominant 
maximum at the point
\begin{equation}
  Q_{xl}=\gamma\sqrt{2N+1} \ \  \text{with} \ \  \gamma\approx2.0.
\label{Q_max}
\end{equation}

We need to consider only small $l$  because $P_{Nl}(Q_{xl})$   is  a monotonically 
decreasing function of $l$, thus, the term $f_{N}-f_{N+l}$ in Eq.~(\ref{rho_xx}) means
the transition should occur at the vicinity of Fermi level, \textit{i. e.}, we have 
$N\approx N_{F}$ where $N_{F}$ is the Landau level index at the Fermi level. 
Therefore Eq. ~(\ref{Q_max}) is equivalent to
\begin{equation}
  \Delta Y_{l}=Q_{xl}l_{B}=\gamma R_{c}\approx2R_{c},
\label{deltaY_max}
\end{equation}
where $R_{c}=\sqrt{2N_{F}+1}l_{B}=l_{B}^{2}\sqrt{2\pi n_{e}}$.
Comparing with Eq.~(\ref{deltaY_fix}), the condition Eq.~(\ref{deltaY_max}) 
leads to
\begin{equation}
 B_{l}=\gamma\frac{\sqrt{2\pi}m^{\ast}}{e^{2}}\frac{1}{\sqrt{n_{e}}}\frac{J_{dc}}{l},
\label{Resonance_condition}
\end{equation}
which explains well the result of Eq. (1). 

From the slopes of the fan diagram in Fig. 2 (b), we obtain $\gamma=1.72, 
1.63, 1.88, 2.05$ for the Hall bars with width $w=200, 100, 50, 20 \ \mu m$ 
respectively. Such values, determinded experimentally, are close enough to 
the theoretical value $\gamma\approx2.0$, indicating the validity of our model 
even in a quantitative sense. Moreover,  we measured the density dependence 
of the oscillation maxima, and plot the results in Fig.~4. It is clearly shown that  
peak positions  $\propto 1/\sqrt{n_{e}}$.

\begin{figure}
\includegraphics{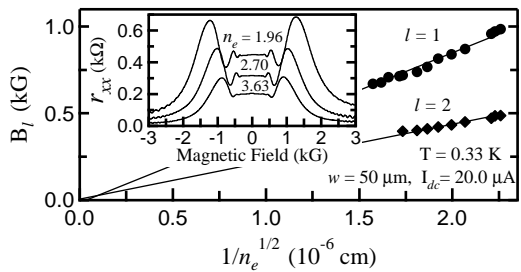}
\caption{\label{fig4} 
The density dependence measurement shows that the maximum positions
are scaled as $1/\sqrt{n_{e}}$. The inset shows measured differential
magnetoresistance traces at selected electron densities (in units of $10^{11}\ cm^{-2}$).}
\end{figure}

The resonance condition Eq.~(\ref{deltaY_max}) can be interpreted 
semiclassically as  following. From Eq.~(\ref{sigma_yy}), the conductivity 
$\sigma_{yy}$ is proportional to the transition rate $W_{\mu\mu^{\prime}}$ 
between two Landau levels near the Fermi level. The transition rate 
drastically goes to zero when $\Delta Y>R_{N}+R_{N^{\prime}}
\approx2R_{c}$ because within this region there is almost no overlap 
between the oscillatory wave functions, as shown in Fig.~3. Thus the 
furthest distance the electron can hop is around 
$(\Delta Y)_{\max}\approx2R_{c}$. Note that $\sigma_{yy}$\ is 
proportional to $\Delta Y$, so naturally a conductivity peak appears at $\Delta 
Y=(\Delta Y)_{\max}\approx2R_{c}$. 

The hopping at a distance $2R_{c}$  along $y$ direction is equivalent to a
momentum transfer along $x$ direction $\Delta k_{x}=2R_{c}l_{B}^{2}=2k_{F}$. 
It is interesting to point out that a similar  momentum transfer mechanism 
has been used to account for the magneto-acoustic phonon  resonance 
of a 2DEG~\cite{Zudov2001}.

In a high-mobility 2DEG, the elastic scatterers for electrons are mainly ionized
impurities in the remote doping layer, residual background ionized impurities
throughout the material, interface roughness, and neutral impurities in the GaAs
well~\cite{Stormer1983}. The remote ionized impurity scattering is long ranged
and in momentum space its potential is exponentially confined into a narrow 
range with a characteristic momentum $q_{s}\sim 1/d_{s}$~\cite{Ando1982}. 
In our samples $d_{s}=70\ nm$, equivalent to a $q_{s}
\sim 0.014\ nm^{-1}$ which is much less than $2k_{F}\approx0.22\ nm^{-1}$; 
therefore the remote ionized impurities are not likely to contribute to the 
oscillations. The other three mechanisms mentioned above are short ranged, 
and therefore in principle could contribute to the oscillations we are discussing. 
The scattering length of the interface roughness and neutral scatters are both 
at atomic scale, so their potential seen by the 2DEG in momentum space are 
all almost constant within the scale of $2k_{F}$.  Our numerical result of 
Eq.~(\ref{deltaY_max}) is modeled on this fact. It is usually assumed that the 
remote ionized impurities are the main sources of scattering for 2DEGs in 
GaAs/AlGaAs heterostructures. However, for high-mobility samples with a wide 
spacer, the residual impurities and interface roughness become 
important~\cite{Saku1996_Umansky1997}. Indeed, according to a theoretical 
study~\cite{Mirlin2001}, the negative magnetoresistance as shown in the 
$I_{dc}=0\ \mu A$ trace of Fig.~1 is a strong evidence 
for the significance of such short-range scatterers in our samples.

In conclusion, we have observed a novel type of magnetoresistance oscillations
in a laterally confined high-mobility 2DEG, which can be attributed to the spatial hopping 
of electrons between tilted Landau levels under a current-induced Hall field. 
Strong short-range scattering is needed for this class of oscillations to occur. 

We comment on possible application of such measurements to the composite
fermion regime in a half-filled Landau level~\cite{Sarma1997}, where the Fermi 
surface effects have been well established but many interesting questions 
remain open. Due to the selection rules of both momentum and energy, the 
cyclotron energy (hence the effective mass) of the CF state could be explored by 
Zener resonance. From the Zener tunneling mechanism point of view, since such oscillations 
require a finite momentum transfer, observation of this effect with CF would 
allow a look into their short-range interactions. Furthermore, in the CF regime 
additional scattering mechanism such as fluctuations of the effective magnetic 
field would be interesting to investigate. 

We benefit from valuable discussions with D. C. Tsui and J. M. Worlock for 
early work in Zener  tunneling in 2DEG. This work was partially supported 
by NSF (C. L. Y., J. Z., and R. R. D.). The work at Sandia was supported by DOE.


\begin{thebibliography}{}
\bibitem{Prange1990} \textit{The Quantum Hall Effect}, edited by R. E. Prange and     S. M. Girvin (Springer-Verlag, New York, 1990).
\bibitem{Ebert_Cage1983} G. Ebert \textit{et al.}, J. Phys. C \textbf{16}, 5441 (1983);
                                            M. E. Cage \textit{et al.}, Phys. Rev. Lett. \textbf{51}, 1374 (1983).
\bibitem{Simon1986} Ch. Simon \textit{et al.}, Phys. Rev. B \textbf{33}, 1190 (1986).
\bibitem{Tsui1983} D. C. Tsui, G. J. Dolan, and A. C. Gossard, 
   Bull. Am. Phys. Soc. \textbf{28}, 365 (1983).
\bibitem{Gerhardts_Winkler1989} R. R. Gerhardts, D. Weiss, and K. v. Klitzing, 
      Phys. Rev. Lett. \textbf{62}, 1173 (1989); R. W. Winkler, J. P. Kotthaus, and K. Ploog, 
       Phys. Rev. Lett. \textbf{62}, 1177 (1989).
\bibitem{Sarma1997} \textit{Perspectives in Quantum Hall Effect}, edited by S. Das Sarma
        and A. Pinczuk (Wiley and Sons, New York, 1997). 
\bibitem{data_analysis} The resonance condition is 
    Eq.~(\ref{Resonance_condition}). From Fourier analysis point of view, 
    it gives an oscillatory resistivity $\rho_{xx}\propto\cos(2\pi l)$, where 
    $l\propto J_{dc}/B$. Accordingly, the voltage $V\propto\rho_{xx}J_{dc}
    \propto J_{dc}\cos(2\pi l)$ and  $r_{xx}\equiv \partial V/\partial I \propto 
    \partial V/\partial J_{dc}$. It is easy to show that $r_{xx}$ has a phase shift
    $\sim 90^{\circ}$ with respect to  $\rho_{xx}$, which can be largely restored
    by $\partial r_{xx}/\partial|B|$.
\bibitem{Kahn1959} A. H. Kahn and H. P. Frederikse, in \textit{Solid State Physics}, 
      edited by  F. Seitz and D. Turnbull (Academic, New York, 1959), Vol \textbf{9}, p. 257.
\bibitem{Calecki1977} D. Calecki, C. Lewiner, and P. Nozieres, J. Phys. (Paris) \textbf{38}, 
        169 (1977). 
\bibitem{Chaubet1995} C. Chaubet, A. Raymond, and D. Dur, 
          Phys. Rev. B \textbf{52}, 11178 (1995). 
\bibitem{Titeica1935} S. Titeica, Ann. Phys. (Leipzig) \textbf{22}, 129 (1935). 
\bibitem{Roth1966} L. M. Roth and P. N. Argyres, in \textit{Semiconductors and 
     Semimetals}, edited by R. K. Willardson and A. C. Beer (Academic, Now York, 1966), 
      Vol. \textbf{1}, p. 159. 
\bibitem{Zudov2001} M. A. Zudov \textit{et al.}, Phys. Rev. Lett.  \textbf{86}, 3614 (2001).
 \bibitem{Stormer1983} H. L. Stormer, Surf. Sci.  \textbf{132}, 519 (1983). 
 \bibitem{Ando1982} T. Ando, J. Phys. Soc. Jpn.  \textbf{51}, 3900 (1982).
 \bibitem{Saku1996_Umansky1997} T. Saku, Y. Horikoshi and Y. Tokura, Jpn. J. Appl. Phys.  
             \textbf{35}, 34 (1996); V. Umansky, R. de-Picciotto, and M. Heiblum, 
              Appl. Phys. Lett.  \textbf{71}, 683 (1997). 
 \bibitem{Mirlin2001} A. D. Mirlin \textit{et al.}, Phys. Rev. Lett.  \textbf{87}, 126805 (2001).
\end{thebibliography}
\end{document}